\documentclass[twocolumn,showpacs,amssymb,aps]{revtex4}
\usepackage{graphicx}

\begin{document}

\title{Ultrarelativistic electron-hole pairing in graphene bilayer}

\author{Yu.\,E. Lozovik}
\email{lozovik@isan.troitsk.ru}
\author{A.\,A. Sokolik}

\affiliation{Institute of Spectroscopy, Russian Academy of Sciences, 142190 Troitsk, Moscow Region, Russia}

\begin{abstract}
We consider ground state of electron-hole graphene bilayer composed of two independently doped graphene layers when a condensate of spatially
separated electron-hole pairs is formed. In the weak coupling regime the pairing affects only conduction band of electron-doped layer and valence
band of hole-doped layer, thus the ground state is similar to ordinary BCS condensate. At strong coupling, an ultrarelativistic character of electron
dynamics reveals and the bands which are remote from Fermi surfaces (valence band of electron-doped layer and conduction band of hole-doped layer)
are also affected by the pairing. The analysis of instability of unpaired state shows that s-wave pairing with band-diagonal condensate structure,
described by two gaps, is preferable. A relative phase of the gaps is fixed, however at weak coupling this fixation diminishes allowing gapped and
soliton-like excitations. The coupled self-consistent gap equations for these two gaps are solved at zero temperature in the constant-gap
approximation and in the approximation of separable potential. It is shown that, if characteristic width of the pairing region is of the order of
magnitude of chemical potential, then the value of the gap in the spectrum is not much different from the BCS estimation. However, if the pairing
region is wider, then the gap value can be much larger and depends exponentially on its energy width.
\end{abstract}


\maketitle

\section{Introduction}
Graphene, being a carbon layer of monoatomic thickness with honeycomb lattice, was fabricated for the first time in 2004
\cite{Novoselov1,Novoselov2}. Large carrier mobility makes graphene promising for possible nano-electronic applications \cite{Geim,Katsnelson1}. Due
to symmetry properties of honeycomb lattice, graphene has unique electron properties: its valence and conduction bands touch at two inequivalent
points of the first Brillouin zone, and near these points (in two ``valleys'') electron effective envelope wave function obeys two-dimen\-sional
Dirac equation for massless particles with the Fermi velocity $v_\mathrm{F}\approx10^6\,\mbox{m/s}\approx c/300$ playing the role of effective
``speed of light''. This leads to a lot of peculiarities of graphene electron properties different from properties of conventional
quasi-two-dimensional electron gas \cite{CastroNeto}.

The possibility to study massless charged fermions in graphene provides a bridge between condensed matter phy\-sics and quantum electrodynamics,
allowing laboratory investigation of relativistic effects which normally require extremal conditions \cite{Katsnelson2}. Thus, it is of interest to
study various coherent phases in graphene and superconductivity in particular. As possible origins of intrinsic superconductivity, phonon- and
plasmon-mediated mechanisms in metal-coated graphene \cite{Uchoa}, electron correlations \cite{Black-Schaffer} and anisotropic electron-electron
scattering near Van Hove singularity \cite{Gonzalez} were proposed.

In electron systems at certain conditions, a condensation of paired electrons and holes as a result of their Coulomb attraction can occur, and this
condensation, like superconductivity, is accompanied by the onset of both an order parameter and the gap in single-particle excitation spectrum of
the system. This problem was first considered in connection with excitonic insulator state \cite{Keldysh}, however transitions between bands of
paired particles in this state lead to order parameter phase fixation and make a superfluidity impossible. On the contrary, in coupled
quasi-two-dimensional quantum wells, spatial separation of electrons and holes strongly suppress the phase fixation and allow superfluid flows, which
appear as persistent electric currents flowing in opposite directions in different layers \cite{Lozovik_Yudson}. Experimental evidences of
pair-condensed state formation were found in course of further research on electron-hole pairing in such or similar systems (see the reviews
\cite{Moskalenko,Timofeev,Butov,Eisenstein}).

The recent success in fabrication of a bilayer graphene system consisting of two parallel electrically independent graphene layers separated by a
nanometer-scale space \cite{Schmidt1}, makes the problem of electron-hole pairing in graphene bilayer to be of high importance. Independent doping of
two graphene layers would allow to achieve equal concentrations of electrons and holes in these layers, whereas spatial interlayer separation $D$
suppress interlayer tunneling and thus electron-hole recombination (moreover, band-structure properties of each layer are unaffected by the presence
of the other one due to negligible interlayer overlap of electron wave functions, in contrast to \emph{bilayer gra\-phene} with Bernal stacking
\cite{McCann}). Such a system was considered first in \cite{Lozovik_Sokolik} in the weak coupling regime, while a possibility to achieve
experimentally both weak- and strong coupling regimes was argued. However, under increase of the coupling strength, a crossover to Bose-Eins\-tein
condensate (BEC) of local pairs, typical for attracting fermion systems (the BCS-BEC crossover \cite{Nozieres}), does not occur in graphene, since
electrons and holes cannot form localized pairs due to absence of gap in graphene spectrum \cite{Note1,Lozovik_Sokolik_JConf}. Instead of a gas of
local pairs, an ``ultrarelativistic'' BCS state should be formed in graphene bilayer at strong coupling. In this state, electrons and holes are
condensed as strongly overlapped Cooper pairs, as in ordinary BCS state, but at the same time an effectively ultrarelativistic nature of fermions
plays an essential role. Unlike the weak coupling regime, where the pairing correlations affect only electrons and holes in narrow regions near their
Fermi surfaces, in the ultrarelativistic regime particles from remote bands (valence band of electron-doped layer and conduction band of hole-doped
one) are also involved in the pairing. Involvement of the remote bands into pairing process is the characteristic feature of ``ultrarelativistic''
BCS state in graphene bilayer in the strong coupling regime. In contrast to BEC of electron-hole pairs, ``ultrarelativistic'' BCS state can possess
rather high critical temperature and anomalous superfluid properties.

Besides the paper \cite{Lozovik_Sokolik}, electron-hole pair condensation in graphene bilayer has been considered also in several other interesting
works \cite{Min,Zhang,Seradjeh,Kharitonov1,Bistritzer,Kharitonov2}. In \cite{Min,Zhang} the problem has been studied in the Hartree-Fock
approximation using unscreened Cou\-lomb potential as a pairing interaction. Estimations \cite{Min,Zhang} of a temperature at which the system turns
into a superfluid state, obtained from numerical calculations, demonstrate that the superfluidity can survive up to room temperatures. However,
taking into account a screening of the interaction can substantially lower the transition temperature down to values of the order of
$10^{-3}\,\mbox{K}$ \cite{Kharitonov1} (see also Refs.~\cite{Bistritzer,Kharitonov2} concerning discussion of a role of self-consistent weakening of
the screening due to appearance of the gap).

Note also the consideration \cite{Kopnin} based upon extension of the BCS theory on cone-shaped band structure, and the work \cite{Aleiner} studying
condensation of electrons and holes with antiparallel spins emerging in graphene due to Zeeman splitting by in-plane magnetic field. Estimations of a
highest possible critical temperature, made in the latter work taking into account the screening of the pairing Coulomb interaction, provide a result
of the order of tens millikelvins.

In discussing the question of quantitative prediction of the critical temperature of the superfluidity in electron-hole graphene bilayer, one should
admit that, in contrast to the weak coupling regime, where applicability of the BCS theory \cite{Lozovik_Sokolik} is doubtless, at strong coupling it
is necessary to take into account a number of factors going beyond BCS approach and capable to change essentially a value of the critical
temperature. These factors are, firstly, multi-band character of the pairing and frequency dependence of the screened pairing interaction, and,
secondly, various correlation effects, missed in the mean-field approximation. Leaving a consideration of the role of dynamical effects and
correlations to a future work, in the present paper we focus our attention on multi-band character of the pairing. We use mean-field approximation,
and the Coulomb potential, statically screened according to the random phase approximation, is assumed as a pairing interaction.

Some estimations of the role of multi-band character of BCS-like pairing of relativistic and ultrarelativistic fermions were made earlier both in
connection with generalization of Nambu-Jona-Lasinio model \cite{Pisarski,Ohsaku1} and for superconductivity in graphene \cite{Kopnin,Ohsaku2} (for a
simplified model of contact pairing potential in the latter case). In the present work, we perform analysis of the problem of electron-hole pairing
in graphene bilayer taking into account a finite range of screened pairing interaction and using physically realistic estimations for an ultraviolet
momentum cutoff in gap equations. By means of electron-hole transformation, the consideration made can be extended to a problem of electron-electron
pairing in a single graphene layer (see also \cite{Lozovik_Ogarkov}).

The paper is organized as follows: the Gor'kov equations and gap equations are derived in Sec.~\ref{part2} and solved approximately at $T=0$ in Sec.~
\ref{part3}. In Sec.~\ref{part4} the instability of unpaired state at finite temperature is analyzed, and in Sec.~\ref{part5} the conclusions are
made.

\section{Description of the paired state}\label{part2}
Near two inequivalent points of the Brillouin zone (two Dirac points) the dynamics of free electrons in graphene is described by the equation
\begin{equation}
v_\mathrm{F}(\vec\sigma\cdot\vec{p})\psi=E\psi,\label{one_Dir}
\end{equation}
where $\vec\sigma=\{\sigma_x,\sigma_y\}$ is a two-dimensional vector composed from Pauli matrices, $\psi$ is a two-component electron wave function,
and its components are physically the envelope wave functions, i.e. the amplitudes of Bloch functions placed on two sublattices of graphene lattice
\cite{Semenoff}. The solutions of (\ref{one_Dir}) are the eigenfunctions $\psi_{\vec{p}\gamma}(\vec{r})=\exp(i\vec{p}\cdot\vec{r})f_{\vec{p}\gamma}$
having a form of a plane wave multiplied by a spinor $f_{\vec{p}\gamma}=\{\exp(-i\varphi/2),\gamma\exp(i\varphi/2)\}^T/\sqrt{2}$, where $\varphi$ is
an azimuthal angle of the momentum vector $\vec{p}$, measured from some fixed direction, $\gamma=\pm1$ is an index of a band, populated by the
electron. The electron is located in the conduction band and its energy (measured from the Dirac point) is $E=v_\mathrm{F}|\vec{p}|$ at $\gamma=+1$,
and in the valence band with $E=-v_\mathrm{F}|\vec{p}|$ at $\gamma=-1$.

To describe an ultrarelativistic condensate, one could use a diagrammatic technique of quantum electrodynamics \cite{Akhiezer} in the limit of
massless electrons, but a modification of non-relativistic diagrammatic technique \cite{Abrikosov} will be used instead, since it reveals a physics
of the multi-band pairing more clearly. For this purpose we introduce the following Matsubara Green functions:
$G^{(\alpha\beta)}_{\gamma_1s_1\gamma_2s_2}(\vec{p},\tau)=-\langle
Tc^{(\alpha)}_{\vec{p}\gamma_1s_1}(\tau)c^{(\beta)+}_{\vec{p}\gamma_2s_2}(0)\rangle$, where $T$ is an ordering operator in imaginary time $\tau$,
$\langle\ldots\rangle$ is an average over the thermodynamic ensemble; $c^{(1)}_{\vec{p}\gamma s}=a^{(1)}_{\vec{p}\gamma s}$, $c^{(2)}_{\vec{p}\gamma
s}=a^{(2)}_{\vec{p}-\gamma s}$, where $a^{(\alpha)}_{\vec{p}\gamma s}$ is the destruction operator for electron with a momentum $\vec{p}$, located in
the layer $\alpha$ ($\alpha=1$ and $2$ mean the electron- and hole-doped layers respectively), having a spin projection and valley denoted jointly by
the index $s$, and populating the band $\gamma$. The switching from $a$ to $c$ operators corresponds to interchanging of valence and conduction bands
in the hole-doped layer. This allows the bands of both kinds of pairing particles, containing their Fermi surfaces, to be denoted by $\gamma=+1$ and
named as ``conduction bands'', while the bands remote from the Fermi surfaces can be denoted by $\gamma=-1$ and named as ``valence bands''. The bare
Green functions are diagonal over all indices and equal to $G^{0(\alpha\alpha)}_{\gamma s\gamma
s}(\vec{p},i\omega_n)=(i\omega_n-\xi^{(\alpha)}_{\vec{p}\gamma})^{-1}$, where
$\xi^{(1)}_{\vec{p}\gamma}=-\xi^{(2)}_{\vec{p}\gamma}\equiv\xi_{p\gamma}=\gamma v_\mathrm{F}|\vec{p}|-\mu$ are the energies of electrons and holes
measured from their chemical potentials (due to electron-hole symmetry of graphene, the Fermi surfaces for electrons and holes coincide at opposite
chemical potentials in two layers).

Thus, two ends of the line, denoting the Green function $G^{(\alpha\beta)}_{\gamma_1s_1\gamma_2s_2}(\vec{p},i\omega_n)$, are attributed by layer
($\alpha,\beta$), band ($\gamma_1,\gamma_2$) and spin-valley $(s_1,s_2)$ indices. Since an intra- and interlayer Coulomb interactions are different,
we also assign the layer indices to each end of the interaction line $V^{(\alpha\beta)}$. As a result of folding over components of spinor electron
wave functions, an additional factor, depending on bands and momentum directions of ingoing and outgoing electrons, should be added to each diagram
vertex. If the electron with the momentum $\vec{p}$ and band index $\gamma$ enters the vertex, and the electron with the momentum $\vec{p'}$ and band
index $\gamma'$ exits, then we attribute the factor
\begin{eqnarray}
\langle f_{\vec{p}\gamma}|f_{\vec{p'}\gamma'}\rangle=\frac12\left\{\exp\frac{i(\varphi-\varphi')}2+
\gamma\gamma'\exp\frac{i(\varphi'-\varphi)}2\right\},\label{ang_fact}
\end{eqnarray}
to it, where $\varphi$ and $\varphi'$ are the azimuthal angles of $\vec{p}$ and $\vec{p'}$ respectively. Calculating an analytical expression
corresponding to a diagram, one should perform summation on layer, band and spin-valley indices in all internal vertices, taking into account that
layer and spin-valley indices are conserved in the vertices, while band indices are not.

\begin{figure}[t]
\begin{center}
\includegraphics[width=0.8\columnwidth]{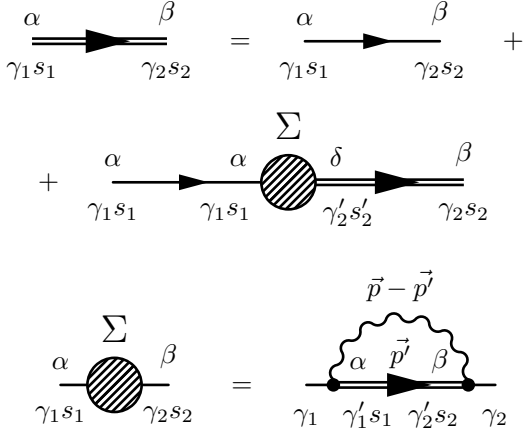}
\end{center}
\caption{\label{fig1}Diagrammatic representation of the Gor'kov (\ref{Gor1}) and self-consistency (\ref{Sc1}) equations.}
\end{figure}

Being guided by the above-stated Feynman rules, we draw the diagrams expressing the Gor'kov equations \cite{Gor'kov}, that describe the pairing in
the mean-field approximation, and also the self-consistency conditions for the self-energies (Fig.~\ref{fig1}). The corresponding analytical
expressions are
\begin{eqnarray}
G^{(\alpha\beta)}_{\gamma_1s_1\gamma_2s_2}(\vec{p},i\omega_n)=G^{0(\alpha\beta)}_{\gamma_1s_1\gamma_2s_2}(\vec{p},i\omega_n)\nonumber\\+
\sum_{\delta\gamma_2's_2'}G^{0(\alpha\alpha)}_{\gamma_1s_1\gamma_1s_1}(\vec{p},i\omega_n)\Sigma^{(\alpha\delta)}_{\gamma_1s_1\gamma_2's_2'}(\vec{p})
G^{(\delta\beta)}_{\gamma_2's_2'\gamma_2s_2}(\vec{p},i\omega_n),\label{Gor1}
\end{eqnarray}
\begin{eqnarray}
\Sigma^{(\alpha\beta)}_{\gamma_1s_1\gamma_2s_2}(\vec{p})=-T\sum_{\omega_n\gamma_1'\gamma_2'}\int\frac{d\vec{p'}}{(2\pi)^2}
V^{(\alpha\beta)}(|\vec{p}-\vec{p'}|)\nonumber\\ \times\langle f_{\vec{p}\gamma_1}|f_{\vec{p'}\gamma_1'}\rangle\langle
f_{\vec{p'}\gamma_2'}|f_{\vec{p}\gamma_2}\rangle G^{(\alpha\beta)}_{\gamma_1's_1\gamma_2's_2}(\vec{p'},i\omega_n),\label{Sc1}
\end{eqnarray}
where $V^{(\alpha\beta)}(q)$ are the Fourier transforms of a statically screened intra- and interlayer electron-electron interaction. Here we do not
take into account frequency dependence of self-energies, assuming that the role of dynamical effects is reduced to some effective restriction of the
pairing region in the momentum space, analogously to restriction of superconducting pairing in metals to the Debye frequency-sized region near the
Fermi surface \cite{BCS}. Since we are interested in the interlayer pairing only, then only the terms of (\ref{Gor1}) with the anomalous
self-energies $\Sigma^{(12)}$ and $\Sigma^{(21)}$ are essential. The terms with the normal self-energies $\Sigma^{(\alpha\alpha)}$, leading to a
renormalization of the Fermi velocity and chemical potential, will be omitted.

The electrons and holes are spatially separated, therefore particles with any spin directions and from any valleys are equally suitable to be paired.
This leads to an arbitrariness of a structure of the anomalous Green functions $G^{(12)}_{\gamma_1s_1\gamma_2s_2}$ and
$G^{(12)}_{\gamma_1s_1\gamma_2s_2}$, playing the role of a condensate wave function, on the indices $s_{1,2}$. In spatially homogeneous case, this
spin-valley structure factorizes in a form of some $(4\times4)$ matrix, denoted by $P$:
$G^{(12)}_{\gamma_1s_1\gamma_2s_2}=P_{s_1s_2}G^{(12)}_{\gamma_1\gamma_2}$,
$G^{(21)}_{\gamma_1s_1\gamma_2s_2}=P^+_{s_1s_2}G^{(21)}_{\gamma_1\gamma_2}$. According to (\ref{Sc1}), the same factorization takes place for the
anomalous self-ener\-gies: $\Sigma^{(12)}_{\gamma_1s_1\gamma_2s_2}=P_{s_1s_2}\Delta_{\gamma_1\gamma_2}$,
$\Sigma^{(21)}_{\gamma_1s_1\gamma_2s_2}=P^+_{s_1s_2}\Delta^+_{\gamma_1\gamma_2}$. Note, that the matrix $P$ should be unitary --- if not, then, as
can be shown, excitations inside the gap appear and the energy of the system increases in comparison with the unitary case (thus, the unitarity of
$P$ means the absence of unpaired particles; see also \cite{Sigrist} in connection with unitarity of a condensate). The normal Green functions are
diagonal on spins and valleys: $G^{(\alpha\alpha)}_{\gamma_1s_1\gamma_2s_2}=\delta_{s_1s_2}G^{(\alpha\alpha)}_{\gamma_1\gamma_2}$.

Since the order parameter has a form of $U(4)=U(1)\times SU(4)$ unitary matrix $P$, fluctuations of both the common phase of the condensate (the
$U(1)$ sector) and of its spin-valley structure (the $SU(4)$ sector) are possible and characterized by the corresponding phase stiffnesses. Moreover,
such topological excitations of the order parameter, as vortices, half-vortices and quarter-vortices can arise (similarly to \cite{Aleiner}, where
the order parameter has a form of $U(2)=U(1)\times SU(2)$ matrix).

For simplicity, we consider the case of the gap, isotropic by an absolute value, i.e. $|\Delta_{\gamma_1\gamma_2}(\vec{p})|$ does not depend on a
direction of $\vec{p}$. At $l$-wave pairing, we have $G^{(12)}_{\gamma_1\gamma_2}(\vec{p},i\omega_n)=e^{il\varphi}F_{\gamma_1\gamma_2}(p,i\omega_n)$,
$G^{(21)}_{\gamma_1\gamma_2}(\vec{p},i\omega_n)=e^{-il\varphi}F^+_{\gamma_1\gamma_2}(p,i\omega_n)$ and
$\Delta_{\gamma_1\gamma_2}(\vec{p})=e^{il\varphi}\Delta_{\gamma_1\gamma_2}(p)$, where $\varphi$ is the azimuthal angle of $\vec{p}$; the normal Green
functions $G^{(\alpha\alpha)}_{\gamma_1\gamma_2}$ are independent on this angle. Using the aforecited notations, write down the systems of Gor'kov
equations (\ref{Gor1}), describing the pairing:
\begin{eqnarray}
\left\{\begin{array}{l}[i\omega_n-\xi^{(1)}_{p\gamma_1}]G^{(11)}_{\gamma_1\gamma_2}-\Delta_{\gamma_1\gamma}F^+_{\gamma\gamma_2}=
\delta_{\gamma_1\gamma_2},\\{}[i\omega_n-\xi^{(2)}_{p\gamma_1}]F^+_{\gamma_1\gamma_2}-\Delta^+_{\gamma_1\gamma}G^{(11)}_{\gamma\gamma_2}=0,
\end{array}\right.\label{Gor2a}\\
\left\{\begin{array}{l}[i\omega_n-\xi^{(2)}_{p\gamma_1}]G^{(22)}_{\gamma_1\gamma_2}-\Delta^+_{\gamma_1\gamma}F_{\gamma\gamma_2}=
\delta_{\gamma_1\gamma_2},\\{}[i\omega_n-\xi^{(1)}_{p\gamma_1}]F_{\gamma_1\gamma_2}-\Delta_{\gamma_1\gamma}G^{(22)}_{\gamma\gamma_2}=0.
\end{array}\right.\label{Gor2b}
\end{eqnarray}
Performing integration on the azimuthal angle of $\vec{p'}$ in (\ref{Sc1}) and taking into account (\ref{ang_fact}), we get the self-consistent gap
equations
\begin{eqnarray}
\Delta_{\gamma_1\gamma_2}(p)=-T\sum_{\omega_n\gamma_1'\gamma_2'}\int\frac{p'\,dp'}{2\pi}\left\{\frac{1+\gamma_1\gamma_2\gamma_1'\gamma_2'}4
\right.\nonumber\\ \times V_l(p,p')+\frac{\gamma_1\gamma_1'}4\,V_{l+1}(p,p')\nonumber\\
\left.+\frac{\gamma_2\gamma_2'}4\,V_{l-1}(p,p')\right\}F_{\gamma_1'\gamma_2'}(p',i\omega_n),\label{Sc2}
\end{eqnarray}
with $l$-wave harmonics of the pairing potential defined:
\begin{eqnarray}
V_l(p,p')=\int\limits_0^{2\pi}\frac{d\varphi}{2\pi}\,V\left(\sqrt{p^2+p'^2-2pp'\cos\varphi}\right)e^{-il\varphi}.\label{harm1}
\end{eqnarray}
The mixing of harmonics of different multipolarity is a manifestation of the ``chirality'' of graphene electrons \cite{Katsnelson2}.

Generally, a matrix structure of the equations (\ref{Gor2a})--(\ref{Sc2}) is self-consistent only for certain forms of the matrix structure of the
order parameter $\Delta_{\gamma_1\gamma_2}$. All possible structures of the order parameter, allowing self-consistent solution for a contact pairing
potential, were found in \cite{Ohsaku1,Ohsaku2}. It was shown, that only scalar and pseudoscalar order parameters has a physical meaning of an
isotropic-gap pairing turning into usual BCS-like pairing in the non-relativistic limit. In our notations, this correspond to a gap matrix,
proportional to the unit matrix: $\Delta_{\gamma_1\gamma_2}\propto\delta_{\gamma_1\gamma_2}$. However, the distinction of our finite-range pairing
potential (\ref{harm1}) from the contact one $V_l(p,p')=V_0\delta_{l0}$ leads to a necessity to modify this solution. We have to assume that the
diagonal elements $\Delta_{\gamma\gamma}$ can differ (similarly to \cite{Pisarski}). The assumption of the band-diagonal pairing is also the simplest
physically, since it corresponds to a pairing of electrons and holes from the bands containing their Fermi surfaces with each other and, at the same
time, to a mutual pairing of electrons and holes from the remote bands. Using $V_{-1}=V_1$, we see from (\ref{Gor2b}) and (\ref{Sc2}) that the
band-diagonal pairing is self-consistent only in the case of $s$-wave pairing, which will be considered henceforth. Additional arguments for a
band-diagonal pairing will be presented in Sec.~\ref{part4}.

The energies of quasiparticle excitations in the case of the band-diagonal pairing are
\begin{eqnarray}
E_\gamma(p)=\sqrt{(v_\mathrm{F}p-\gamma\mu)^2+|\Delta_{\gamma\gamma}(p)|^2},\label{E_pm}
\end{eqnarray}
and the anomalous Green functions are
\begin{eqnarray}
F_{\gamma\gamma}(p,i\omega_n)=-\frac{\Delta_{\gamma\gamma}(p)}{\omega_n^2+E_\gamma^2(p)}.\label{F}
\end{eqnarray}
Substituting (\ref{F}) into (\ref{Sc2}) and summing up over Matsubara frequencies $\omega_n=(2n+1)\pi T$, we get the following system of gap
equations:
\begin{eqnarray}
\left\{\begin{array}{r}\displaystyle\Delta_{++}(p)=\int\frac{p'\,dp'}{2\pi}\left\{V_a(p,p')\frac{\Delta_{++}(p')}{2E_+(p')}\right.\\
\displaystyle\times\tanh\frac{E_+(p')}{2T}\left.+V_b(p,p')\frac{\Delta_{--}(p')}{2E_-(p')}\tanh\frac{E_-(p')}{2T}\right\},\\\\
\displaystyle\Delta_{--}(p)=\int\frac{p'\,dp'}{2\pi}\left\{V_b(p,p')\frac{\Delta_{++}(p')}{2E_+(p')}\right.\\
\displaystyle\times\tanh\frac{E_+(p')}{2T}\left.+V_a(p,p')\frac{\Delta_{--}(p')}{2E_-(p')}\tanh\frac{E_-(p')}{2T}\right\},\end{array}\right.
\label{Sc3}
\end{eqnarray}
where $V_a=(V_0+V_1)/2$, $V_b=(V_0-V_1)/2$. A nonzero $V_b$ leads to a mixing of $\Delta_{++}$ and $\Delta_{--}$ in (\ref{Sc3}) and hence to the
fixation of the relative phase of these two gaps.

The system (\ref{Sc3}) differs in form from the multi-band gap equation of \cite{Kopnin} since the latter is obtained by a simple extension of BCS
approach on cone-like band structure and misses the electrons chirality. In the limit of the contact pairing interaction, when $V_a=V_b$, this
equation coincides with (\ref{Sc3}).

\section{Gap at zero temperature}\label{part3}

\subsection{Pairing interaction}
A solution of (\ref{Sc3}) is determined by a pairing potential $V(q)$, which can be found taking into account two circumstances. Firstly, a behavior
of the system is governed by two dimensionless parameters \cite{Lozovik_Sokolik}: by the ratio of characteristic Coulomb and quantum kinetic energies
\begin{eqnarray}
r_\mathrm{s}=\frac{e^2}{\varepsilon\hbar v_\mathrm{F}}\approx\frac{2.19}\varepsilon\nonumber
\end{eqnarray}
and by the dimensionless interlayer distance $d=p_\mathrm{F}D/\hbar$, where $p_\mathrm{F}=\mu/v_\mathrm{F}$ is the Fermi momentum (the system of
coupled quantum wells is also governed by the same parameters, but with different $r_\mathrm{s}$ \cite{DePalo}; the local pair regime is
characterized by only one dimensionless density parameter \cite{Lozovik_Yudson,Lozovik_Kurbakov}). Secondly, a large number $N$ of fermionic flavors
allows using $1/N$-expansion and thus validates the random-phase approximation for a screened potential (see \cite{Apenko} and references therein).
In graphene bilayer, at $d\ll1$, the number of flavors $N=2\,(\mbox{spin projections})\times2\,(\mbox{valleys})\times2\,(\mbox{layers})=8$, therefore
the random-phase approximation is valid even at large $r_\mathrm{s}$ \cite{Kharitonov1,Kharitonov2}. The dynamically screened interlayer
electron-electron interaction in this approximation is \cite{Lozovik_Yudson,DasSarma_Madhukar}
\begin{eqnarray}
V(q,\omega)=\frac{v_qe^{-qD}}{1-2v_q\Pi(q,\omega)+v^2_q\Pi^2(q,\omega)[1-e^{-2qD}]},\label{V1}
\end{eqnarray}
where $v_q=2\pi e^2/\varepsilon q$ is a bare Coulomb interaction, $\varepsilon$ is a dielectric permittivity of a medium, surrounding graphene
layers, and $\Pi(q,\omega)$ is a polarization operator of graphene (it is the same for both layers due to the particle-hole symmetry).

The cutoff energy $w$, specifying half-width of a pairing region around the Fermi surface, can be estimated as a characteristic frequency of the
lower branch of plasma oscillations in the bilayer. This frequency, found as the first zero of denominator of (\ref{V1}), gives the region of
frequencies where the pairing potential is attractive \cite{Lozovik_Sokolik}:
\begin{eqnarray}
w=\mu\times\left\{\begin{array}{rcl}\sqrt{8r_\mathrm{s}/d},&\mbox{if}&r_\mathrm{s}d\gg1,\\
8r_\mathrm{s},&\mbox{if}&r_\mathrm{s}d\ll1,\;r_\mathrm{s}\ll1,\\2,&\mbox{if}&r_\mathrm{s}d\ll1,\;r_\mathrm{s}\sim1.\end{array}\right.\label{w}
\end{eqnarray}

In the weak coupling regime, at small $r_\mathrm{s}$ or large $d$, we have $w\ll\mu$, and the pairing does not involve particles from the bands,
remote from the Fermi surfaces, so we suppose $\Delta_{--}=0$. Therefore, at weak coupling only the first equation of the system (\ref{Sc3}) remains:
\begin{eqnarray}
\Delta_{++}(p)=\int\frac{p'\,dp'}{2\pi}V_a(p,p')\frac{\Delta_{++}(p')}{2E_+(p')}\tanh\frac{E_+(p')}{2T}.\label{Sc4}
\end{eqnarray}
Due to electron chirality in graphene, this equation involves the half-sum $V_a=(V_0+V_1)/2$ of $s$- and $p$-wave harmonics of the pairing potential
even at $s$-wave pairing instead of only the $s$-wave part involved in the ordinary BCS gap equation \cite{BCS}.

Since $w\ll\mu$, we can suppose the gap to be constant if the pairing region, in the spirit of the BCS approach,
$\Delta_{++}(p)=\Delta\Theta(w-|v_\mathrm{F}p-\mu|)$, where $\Theta(x)=0$ at $x<0$, $\Theta(x)=1$ at $x\geq0$, while the potential $V_a(p,p')$ can be
replaced by its value $V_a(p_\mathrm{F},p_\mathrm{F})$ on the Fermi surface. At $T=0$ we obtain the asymptotical solution of (\ref{Sc4}) in the
BCS-like form \cite{Lozovik_Sokolik}:
\begin{eqnarray}
\Delta=2w\exp\left\{-\frac1{\lambda_a}\right\},\label{Delta1}
\end{eqnarray}
where $\lambda_a=\mathcal{N}V_a(p_\mathrm{F},p_\mathrm{F})$ is the dimensionless constant of intraband coupling, $\mathcal{N}=\mu/2\pi \hbar
v_\mathrm{F}^2$ is the density of states at the Fermi level. The most favorable case for the pairing is $d\ll1$, when the interlayer distance is a
smallest characteristic distance in the system; in this case, using the long-wavelength asymptotics of the static polarization operator
\cite{Wuncsh,Hwang} $\Pi(q,0)\approx-4\mathcal{N}$, the potential (\ref{V1}) takes a simple form:
\begin{eqnarray}
\mathcal{N}V(q,0)=\frac{r_\mathrm{s}}{\tilde{q}+8r_\mathrm{s}}.\label{V3}
\end{eqnarray}
Integrating (\ref{V3}) on angle according to (\ref{harm1}), we get
\begin{eqnarray}
\lambda_a=\frac{r_\mathrm{s}}\pi\left\{2\pi r_\mathrm{s}-1-\sqrt{16r_\mathrm{s}^2-1}\arccos\frac1{4r_\mathrm{s}}\right\}.\label{lambda_a}
\end{eqnarray}
The expressions (\ref{Delta1}) and (\ref{lambda_a}) determine the gap estimation in one-band pairing regime at weak coupling, considered in
\cite{Lozovik_Sokolik}. This estimation gives $\sim10^{-7}\mu$, in agreement with \cite{Kharitonov1}. However, as will be shown below, the gap in a
multi-band pairing regime at strong coupling can be much larger.

Further, for considering the multi-band pairing, we use two models, allowing to reduce the integral gap equations (\ref{Sc3}) to algebraic equations.
These are a constant gap approximation and a separable potential approximation.

\subsection{Constant gap approximation}
Suppose that the gap functions $\Delta_{\gamma\gamma}(p)$ are constant in the pairing region with the energy half-width $w$, partly occupying the
remote bands: $\Delta_{++}(p)=\Delta_+\Theta(w+\mu-v_\mathrm{F}p)$, $\Delta_{--}(p)=\Delta_-\Theta(w-\mu-v_\mathrm{F}p$) (assume $\Delta_+$ and
$\Delta_-$ to be real and positive). Furthermore, replacing the potentials $V_{a,b}$ by their values at the Fermi level, we reduce the system
(\ref{Sc3}) to
\begin{eqnarray}
\left\{\begin{array}{l}\Delta_+=\lambda_aA_+\Delta_++\lambda_bA_-\Delta_-,\\
\Delta_-=\lambda_bA_+\Delta_++\lambda_aA_-\Delta_-,\end{array}\right.\label{Sc5}
\end{eqnarray}
\begin{eqnarray}
A_\pm=\int\limits_0^{w/\mu\pm1}\frac{x\,dx}{2\sqrt{(x\mp1)^2+(\Delta_\pm/\mu)^2}},\label{A1}
\end{eqnarray}
where $\lambda_b=\mathcal{N}V_b(p_\mathrm{F},p_\mathrm{F})$ is a dimensionless constant of interband coupling. Similarly to (\ref{lambda_a}), at
$d\ll1$ we have
\begin{eqnarray}
\lambda_b=\frac{r_\mathrm{s}}\pi\left\{-2\pi
r_\mathrm{s}+1+\frac{16r_\mathrm{s}^2}{\sqrt{16r_\mathrm{s}^2-1}}\arccos\frac1{4r_\mathrm{s}}\right\}.\label{lambda_b}
\end{eqnarray}

\begin{figure}[t]
\begin{center}
\includegraphics[width=0.9\columnwidth]{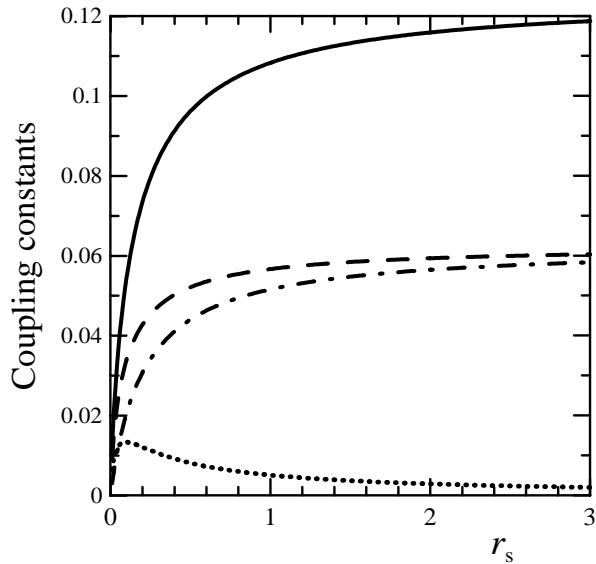}
\end{center}
\caption{\label{fig2}The dimensionless coupling constants at $d\ll1$ as functions of $r_\mathrm{s}$: $\lambda_0$ (solid line), $\lambda_1$ (dotted
line), $\lambda_a$ (dashed line), $\lambda_b$ (dash-dotted line).}
\end{figure}

According to their involvement in (\ref{Sc5}), the constants $\lambda_a$ and $\lambda_b$ are called intra- and interband coupling constants
respectively. By definition, their sum $\lambda_0=\lambda_a+\lambda_b$ is a dimensionless $s$-wave harmonic of the pairing potential at the Fermi
level, whereas the difference $\lambda_1=\lambda_a-\lambda_b$ is a $p$-wave harmonic. When the range of the pairing potential is small (this occurs
at large $r_\mathrm{s}$), the $p$-wave harmonic tends to zero and thus $\lambda_a$ and $\lambda_b$ approach each other. Indeed, at $d\ll1$,
$r_\mathrm{s}\rightarrow\infty$ (\ref{lambda_a}), (\ref{lambda_b}) give $\lambda_0\approx1/8$, $\lambda_1\approx1/48\pi r_\mathrm{s}$, thus
$\lambda_{a,b}\rightarrow1/16$. On the contrary, when the range of the pairing potential is large (this is the case of small $r_\mathrm{s}$), $s$-
and $p$-wave harmonics are close by the order of magnitude, thus one cannot neglect a difference between $\lambda_a$ and $\lambda_b$. Indeed, at
$d\ll1$, $r_\mathrm{s}\ll1$ the expressions (\ref{lambda_a}), (\ref{lambda_b}) give $\lambda_a\approx (r_\mathrm{s}/\pi)(-1-\ln2r_\mathrm{s})$,
$\lambda_b\approx r_\mathrm{s}/\pi$. The plots of $\lambda_0$, $\lambda_1$, $\lambda_a$ and $\lambda_b$ as functions of $r_\mathrm{s}$ at $d\ll1$,
presented in Fig.~\ref{fig2}, confirm these observations.

In asymptotical case of small gaps $\Delta_\pm\ll\mu$, we get from (\ref{A1})
\begin{eqnarray}
A_+\approx\frac{w-\mu}{2\mu}+\ln\frac{2\sqrt{w\mu}}{\Delta_+},\quad A_-\approx\frac{w-\mu}{2\mu}+\ln\sqrt{\frac\mu{w}}.\label{A2}
\end{eqnarray}
Then, substituting (\ref{A2}) into (\ref{Sc5}), we obtain the asymptotical expressions for the gaps:
\begin{eqnarray}
\Delta_+=2\mu\exp\left\{-\left(\frac1{\lambda_a}-\frac{w}\mu+1+\frac{\lambda_a^2-\lambda_b^2}{\lambda_a}\right.\right.\nonumber\\
\left.\left.\times\left[\left(\frac{w}\mu-1\right)^2-\left(\ln\frac{w}\mu\right)^2\right]\right)\right.\nonumber\\
\left.\times\left(1-2\frac{\lambda_a^2-\lambda_b^2}{\lambda_a}\left[\frac{w}\mu-1-\ln\frac{w}\mu\right]\right)^{-1}\right\},\label{Delta2}
\end{eqnarray}
\begin{eqnarray}
\Delta_-=\Delta_+\lambda_b\frac{\frac{w}\mu-1+2\ln\frac{2\sqrt{w\mu}}{\Delta_+}}{2-\lambda_a\left(\frac{w}\mu-1-\ln\frac{w}\mu\right)}.\label{Delta3}
\end{eqnarray}
These expressions differ in form from the BCS result (\ref{Delta1}). If we neglect difference between intra- and interband coupling constants and set
$\lambda_a=\lambda_b$ (this corresponds to a contact pairing potential), then (\ref{Delta2}) is reduced to the asymptotical formula (6) from
\cite{Kopnin}:
\begin{eqnarray}
\Delta_+=2\mu\exp\left\{-\frac2{\lambda_0}+\frac{w}\mu-1\right\}.\label{Delta4}
\end{eqnarray}

\begin{figure}[t]
\begin{center}
\includegraphics[width=0.9\columnwidth]{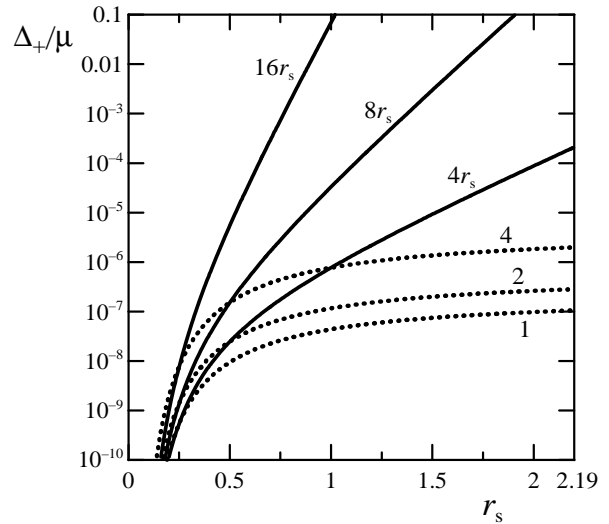}
\end{center}
\caption{\label{fig3}The conduction-band gap $\Delta_+$, normalized by the chemical potential $\mu$, as a function of $r_\mathrm{s}$ at $T=0$,
$d\ll1$ and various ratios of $w/\mu$, indicated near corresponding curves.}
\end{figure}

Thus, in our model the gaps $\Delta_\pm$ at $T=0$ depend on $d$, $r_\mathrm{s}$ and $w$. A rough estimation of the pairing region half-width at
$d\ll1$ is $w=\max(2\mu,8\mu r_\mathrm{s})$, according to (\ref{w}). However, this estimation is only by the order of magnitude, so we consider
various values of $w$ in vicinity of $w=8\mu r_\mathrm{s}$ and $w=2\mu$ for numerical calculations. Fig.~\ref{fig3} shows the conduction-band gap
values $\Delta_+$ obtained by numerical solution of the system (\ref{Sc5}) as functions of $r_\mathrm{s}$ at $d\ll1$ and various ratios of $w/\mu$
(the maximum value of $r_\mathrm{s}\approx2.19/\varepsilon$ is 2.19 in the case when both graphene layers are suspended in vacuum). We see that
$\Delta_+$, firstly, depends on $w/\mu$ approximately exponentially and, secondly, quickly tends to a constant asymptotic at $w=\mathrm{const}$ and
large $r_\mathrm{s}$. This can be easily understood using the approximate formula (\ref{Delta4}) and the plot of $\lambda_0$ from Fig.~\ref{fig2}: on
increase of $r_\mathrm{s}$ at $w=\mathrm{const}$, the exponent of (\ref{Delta4}) tends to the limiting value $w/\mu-17$. Therefore, at large $w$ the
gap $\Delta_+$ depends on $w$ exponentially, in contrast to the BCS result (\ref{Delta1}), where $w$ appears only as a pre-exponential factor.

Nevertheless, according to (\ref{w}), the estimations $w=8\mu r_\mathrm{s}$ and $w=4\mu$ are valid only at $r_\mathrm{s}\ll1$ and $r_\mathrm{s}\sim1$
respectively, so the appropriate values of $\Delta_+$ are given in Fig.~\ref{fig3} by solid lines at small $r_\mathrm{s}$, and by dotted lines at
large enough $r_\mathrm{s}$. Thus, in the model used the ratio $w/\mu$ reach 2 by the order of magnitude at maximum, and the maximum value of
$\Delta_+$ is $(10^{-7}-10^{-6})\mu$. The gap value in the multi-band pairing regime in the model of constant gaps exceeds the BCS estimation only if
$w\gg\mu$, which is not the case for dynamically-screened Coulomb interaction. However, as will be shown below, going beyond the approximation
$\Delta_\pm=\mathrm{const}$ can rise the gaps considerably owing to integration over wide region of momentums in the gap equations.

\begin{figure}[t]
\begin{center}
\includegraphics[width=0.9\columnwidth]{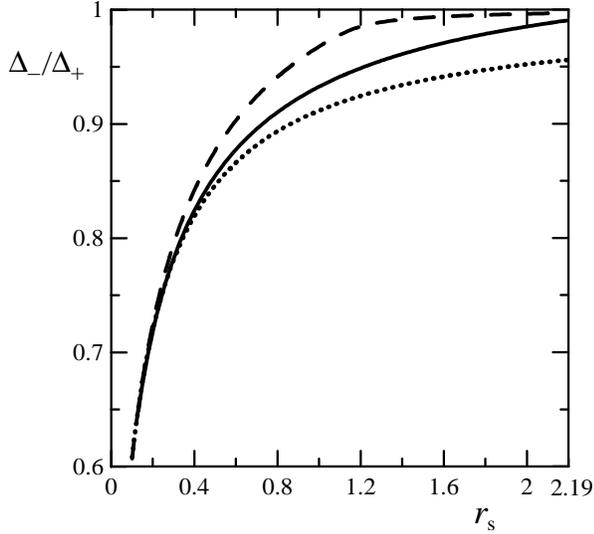}
\end{center}
\caption{\label{fig4}The ratio of the valence-band gap $\Delta_-$ to the conduction-band gap $\Delta_+$ as a function of $r_\mathrm{s}$ at $T=0$,
$d\ll1$ and various ratios of $w/\mu r_\mathrm{s}$: $16r_\mathrm{s}$ (dashed line), $8r_\mathrm{s}$ (solid line), 2 (dotted line).}
\end{figure}

As for the ratio of valence- to conduction-band gaps, the difference of these gaps $\Delta_+$ and $\Delta_-$ is as stronger, as considerably the
constants $\lambda_a$ and $\lambda_b$ differ, as follows from (\ref{Sc5}). The difference between $\lambda_a$ and $\lambda_b$ is large at small
$r_\mathrm{s}$ and small at large $r_\mathrm{s}$, as noted above in the discussion of Fig.~\ref{fig2}. Fig.~\ref{fig4} shows the ratio
$\Delta_-/\Delta_+$ at $T=0$, $d\ll1$ and various ratios $w/\mu$, as functions of $r_\mathrm{s}$. It is seen, that $\Delta_-/\Delta_+$ depends rather
weakly on $w/\mu$ and, in agreement with above argumentation, increases and tends to unity when $r_\mathrm{s}$ increases \cite{Note2}. At
$r_\mathrm{s}>1$, the ratio $\Delta_-/\Delta_+$ is very close to unity, and this means that the pairings in conduction and valence bands are of
equally importance. The multi-band character of the pairing can lead to a number of unusual superfluid properties of the system. Some results
concerning this problem were obtained in \cite{Seradjeh,Milovanovic}, where properties of vortices and zero-energy modes inside these vortices were
studied in the approximation of the contact pairing potential in multi-band regime.

In our case, at weak coupling, when $\lambda_b\ll\lambda_a$, the fixation of the relative phase of $\Delta_+$ and $\Delta_-$ becomes weak, and in the
limit $\lambda_b=0$ the phase fixation is absent, since the equations in the system (\ref{Sc5}) decouple. Therefore at small $r_\mathrm{s}$ gapped
and soliton-like excitations, corresponding to oscillations of the relative phase and being solutions of the sine-Gordon equation, can arise
(analogously to \cite{Lozovik_Poushnov}, see also references therein).

\subsection{Approximation of separable potential}
Let now try to go beyond the approximation of a constant gap and take into account dependence of the functions $\Delta_\pm(p)$ on momentum. For this
purpose, keep only a separable parts of the harmonics $V_0(p,p')$ and $V_1(p,p')$ of the statically screened potential (\ref{V3}) (in the same way,
as in \cite{Khodel}):
\begin{eqnarray}
V_l^\mathrm{(sep)}(p,p')=\frac{V_l(p,p_\mathrm{F})V_l(p_\mathrm{F},p')}{V_l(p_\mathrm{F},p_\mathrm{F})}.\label{V4}
\end{eqnarray}
The separable part $V_l^\mathrm{(sep)}(p,p')$ is close to the original potential $V_l(p,p')$ when at least one of the momentums $p$ and $p'$ lies in
a region, which is close to the Fermi surface and provides the major contribution to the integrals in the gap equations (\ref{Sc3}). Thus we can
suppose that the replacement of $V_l(p,p')$ on its separable part $V_l^\mathrm{(sep)}(p,p')$ will not lead to essential mistakes in defining the gap
functions $\Delta_\pm(p)$. A difference between $V_l^\mathrm{(sep)}(p,p')$ and $V_l(p,p')$ can be large if the both momentums $p$ and $p'$ are far
from the Fermi surface, however, as one can see, for the statically screened potential in the form of (\ref{V3}) the condition
$V_l^\mathrm{(sep)}(p,p')<V_l(p,p')$ at any momentums applies, so the replacement of the potential on its separable part can lead only to
underestimation of gap values.

Substituting (\ref{V4}) into (\ref{Sc3}), we see that the functions $\Delta_\pm(p)$ can be represented in the form
\begin{eqnarray}
\begin{array}{l}
\Delta_+(p)=\tilde\Delta_+v_a(p)+\tilde\Delta_-v_b(p),\\
\Delta_-(p)=\tilde\Delta_+v_b(p)+\tilde\Delta_-v_a(p),
\end{array}\label{Delta5}
\end{eqnarray}
where
\begin{eqnarray}
v_{a,b}(p)=\frac{v_0(p)\pm v_1(p)}2,\quad v_l(p)=\frac{V_l(p,p_\mathrm{F})}{V_l(p_\mathrm{F},p_\mathrm{F})}\label{V5}
\end{eqnarray}
(note that $\Delta_\pm(p_\mathrm{F})=\tilde\Delta_\pm$, so the gap in the excitation spectrum equals approximately to $\tilde\Delta_+$). As a result
of the substitutions (\ref{V4})--(\ref{Delta5}), the system (\ref{Sc3}) of the integral equations is reduced to a system of two algebraic equations
with respect to the two \emph{gap parameters} $\tilde\Delta_+$ and $\tilde\Delta_-$:
\begin{eqnarray}
\left\{\begin{array}{l}\tilde\Delta_+=\lambda_aA\tilde\Delta_++\lambda_bC\tilde\Delta_++\lambda_aC\tilde\Delta_-+\lambda_bB\tilde\Delta_-,\\
\tilde\Delta_-=\lambda_aC\tilde\Delta_++\lambda_bA\tilde\Delta_++\lambda_aB\tilde\Delta_-+\lambda_bC\tilde\Delta_-,\end{array}\right.\label{Sc6}
\end{eqnarray}
and the coefficients of these equations
\begin{eqnarray}
A=\tilde{U}_{aa}^++U_{bb}^-,\quad B=U_{bb}^++U_{aa}^-,\quad C=U_{ab}^++U_{ab}^-,\nonumber
\end{eqnarray}
\begin{eqnarray}
U_{ij}^\pm=\frac{v_\mathrm{F}}{p_\mathrm{F}}\int\limits_0^\infty\frac{v_i(p)v_j(p)p\,dp}{2\sqrt{(v_\mathrm{F}p\mp\mu)^2+\Delta_\pm^2(p)}}\label{U1}
\end{eqnarray}
are functions of $\tilde\Delta_\pm$ themselves, as long as $\Delta_\pm(p)$ in the integrals are defined by (\ref{Delta5}). At large $p$ the functions
$v_{a,b}(p)$ behave as $r_\mathrm{s}/2\lambda_0p$, so the integrals (\ref{U1}) converge and we do not need any momentum cutoffs.

In the limit $\tilde\Delta_\pm\rightarrow0$ only $U_{aa}^+$ from the all integrals (\ref{U1}) is logarithmically singular and diverges as
$\ln(\mu/\tilde\Delta_+)$, whereas all the other integrals converge to finite limits. For both analytical estimations and numerical calculations, it
is convenient to isolate this singularity as follows: $U_{aa}^+=\ln(2\mu/\tilde\Delta_+)+\tilde{U}_{aa}^+$ and analogously
$A=\ln(2\mu/\tilde\Delta_+)+\tilde{A}$. Denoting the limits of $\tilde{A}$, $B$ and $C$ at $\tilde\Delta_\pm\rightarrow0$ as $\tilde{A}_0$, $B_0$ and
$C_0$ respectively, we can derive an asymptotical solution of the system (\ref{Sc6}) at small $\tilde\Delta_\pm$:
\begin{eqnarray}
\tilde\Delta_+=2\mu\exp\left\{-\left[\vphantom{\lambda_a^2}1-\lambda_a(\tilde{A}_0+B_0)-2\lambda_bC_0\right.\right.\nonumber\\
\left.\left.\left.+(\lambda_a^2-\lambda_b^2)(\tilde{A}_0B_0-C_0^2)\right]\right/
\left[\lambda_a-(\lambda_a^2-\lambda_b^2)B_0\right]\right\},\label{Delta6}
\end{eqnarray}
\begin{eqnarray}
\tilde\Delta_-=\tilde\Delta_+\frac{\lambda_b+(\lambda_a^2-\lambda_b^2)C_0}{\lambda_a-(\lambda_a^2-\lambda_b^2)B_0}.\label{Delta7}
\end{eqnarray}

\begin{figure}[t]
\begin{center}
\includegraphics[width=0.9\columnwidth]{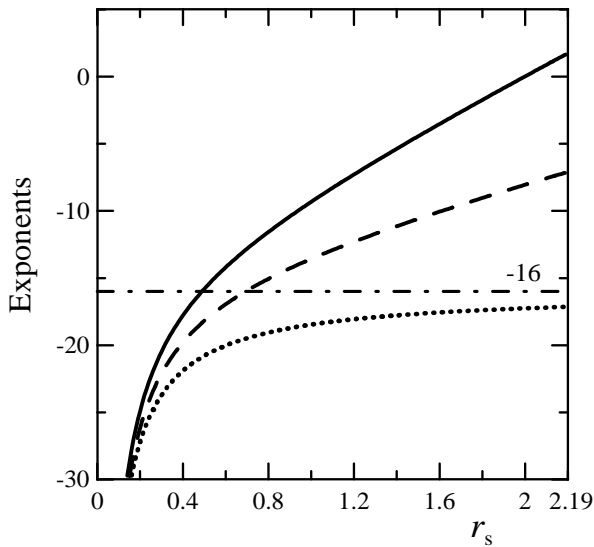}
\end{center}
\caption{\label{fig5}The exponent $a$ in the approximate formula $\tilde\Delta_+=2\mu\exp(a)$ for the gap parameter $\tilde\Delta_+$ as a function of
$r_\mathrm{s}$. Solid line: the result in a framework of the separable potential approximation according to (\ref{Delta8}). Dotted line: the same,
but with neglecting the contribution of the valence band, according to (\ref{Delta9}). Dashed line: the curve $a=-2/\lambda_0$, corresponding to BCS
limit (the dash-dotted line shows the asymptotic $a=-16$ of this curve at large  $r_\mathrm{s}$).}
\end{figure}

\begin{figure}[t]
\begin{center}
\includegraphics[width=0.9\columnwidth]{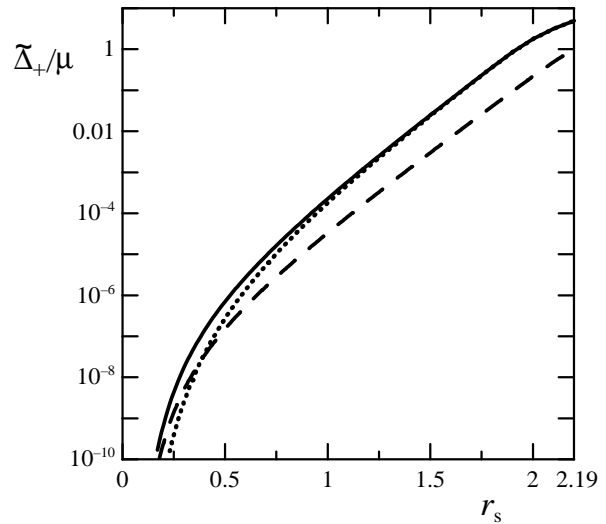}
\end{center}
\caption{\label{fig6}Solid line: the gap parameter $\tilde\Delta_+$, normalized on the chemical potential $\mu$, as a function of $r_\mathrm{s}$
calculated in the separable potential approximation. Dotted line: the result obtained in the approximation $\lambda_a=\lambda_b=\lambda_0/2$, dashed
line: the result for $\Delta_+$ in the approximation $\Delta_\pm=\mathrm{const}$ at $w=8\mu r_\mathrm{s}$ (taken from Fig.~\ref{fig3}).}
\end{figure}

Assuming for simplicity $\lambda_a\approx\lambda_b\approx\lambda_0/2$ (validity of this assumption is confirmed by Fig.~\ref{fig2}), we get
\begin{eqnarray}
\tilde\Delta_+=2\mu\exp\left\{-\frac2{\lambda_0}+\tilde{A}_0+B_0+2C_0\right\}.\label{Delta8}
\end{eqnarray}
The result obtained is similar to (\ref{Delta4}) and differs from the BCS result (\ref{Delta1}) by a presence of additional terms in the exponent;
these terms are positive anywhere except a small region at $r_\mathrm{s}\ll1$. It is interesting to note, that, according to (\ref{U1}), each of
these terms can be divided into two parts corresponding to contributions from conduction and valence bands. Excluding the valence-band contributions
$U_{ij}^-$ from the exponent, we get an expression for a gap, corresponding to one-band model, which takes into account only the conduction band:
\begin{eqnarray}
\tilde\Delta_+^\mathrm{(ob)}=2\mu\exp\left\{-\frac2{\lambda_0}+\tilde{U}_{aa}^++U_{bb}^++2U_{bb}^+\right\}.\label{Delta9}
\end{eqnarray}

In such a way, one part of a positive contribution to the exponent of (\ref{Delta8}) comes from the valence band, whereas the other part comes from
the conduction band and results from smooth dependence of $\Delta_\pm(p)$ and $V_l^\mathrm{(sep)}(p,p')$ on momentums in a wide region.
Fig.~\ref{fig5} presents the exponents $a$ in the approximate formula $\tilde\Delta_+=2\mu\exp(a)$ for the gap parameter as functions of
$r_\mathrm{s}$ in three cases: taking into account the both valence and conduction band contributions, as corresponds to the formula (\ref{Delta8}),
taking into account only conduction band, according to (\ref{Delta9}), and the BCS limit when the exponent is $a=-2/\lambda_0$. We see that, at large
enough $r_\mathrm{s}$ the additional contributions to the exponent from the valence and conduction bands are numerically large, approximately equal
to each other and grow linearly with $r_\mathrm{s}$.

The corresponding results for the gap parameter $\tilde\Delta_+$ are presented in Fig.~\ref{fig6}. These results are obtained by numerical solving of
the system (\ref{Sc6}) in two cases: with taking into account a difference between the coupling constants $\lambda_{a,b}$ or with neglecting this
difference. The difference between two these curves becomes unessential at large $r_\mathrm{s}$, in accordance with decrease of $\lambda_1$ relative
to $\lambda_0$ (see Fig.~\ref{fig2}). Due to dependence of both the gap functions $\Delta_\pm(p)$ and potentials $V_l^\mathrm{(sep)}(p,p')$ on
momentums, the gap $\tilde\Delta_+$ can reach large values comparable to $\mu$; this result demonstrates a striking difference from the BCS result
$\tilde\Delta_+\sim(10^{-7}-10^{-6})\mu$.

For comparison, the curve of $\Delta_+$ in the approximation $\Delta_\pm=\mathrm{const}$ at $w=8\mu r_\mathrm{s}$, analogous to one shown in
Fig.~\ref{fig3}, is also shown in Fig.~\ref{fig6}. A good agreement this curve demonstrates with results obtained within the separable potential
approximation can be readily understood from the expression (\ref{V3}) for the statically screened potential: the characteristic size of a momentum
region providing a major contribution to the gap equations, is $8p_\mathrm{F}r_\mathrm{s}$, which acts similarly to restricting the pairing region to
an energy half-width of $w\sim8\mu r_\mathrm{s}$.

The ratio of the gap parameters $\tilde\Delta_-/\tilde\Delta_+$ obtained by numerical calculations is close both qualitatively and quantitatively to
the results presented in Fig.~\ref{fig4}.

\section{Instability of normal state}\label{part4}
\begin{figure}[t]
\begin{center}
\includegraphics[width=0.9\columnwidth]{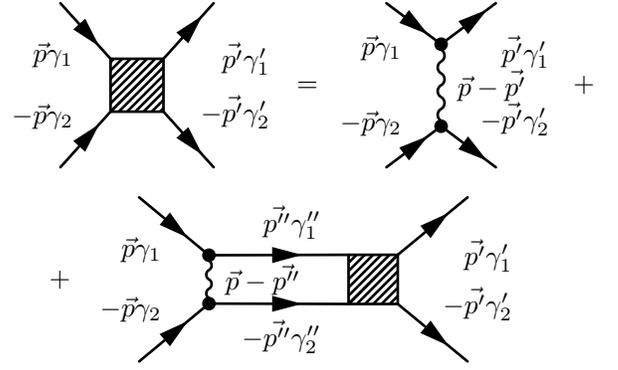}
\end{center}
\caption{\label{fig7}Diagrammatic representation of the Bethe-Salpeter equation (\ref{BS1}).}
\end{figure}

Studying an instability of a normal, unpaired state of the system allows to determine a mean-field critical temperature $T_\mathrm{c}$ of the system
transition into a superfluid state. This mean-field $T_\mathrm{c}$ provides an upper limit for a Kosterlitz-Thouless transition temperature in
two-dimensional system \cite{Kosterlitz}. Consider a vertex part of interlayer electron-hole scattering in the ladder approximation, obeying the
Bethe-Salpeter equation. On decrease of a temperature down to the critical value $T_\mathrm{c}$, a pole appears in the vertex part as a function of a
total energy of scattered particles. The Bethe-Salpeter equation for electron-hole pair at rest has a diagrammatic form shown in Fig.~\ref{fig7}, and
the corresponding analytical expression is
\begin{eqnarray}
\Gamma_{\gamma_1\gamma_2\gamma_1'\gamma_2'}(\vec{p},\vec{p'},E)=V(|\vec{p}-\vec{p'}|)\langle f_{\vec{p}\gamma_1}|f_{\vec{p'}\gamma_1'}\rangle\langle
f_{\vec{p}\gamma_2}|f_{\vec{p'}\gamma_2'}\rangle\nonumber\\-T\int\frac{d\vec{p''}}{(2\pi)^2}V(|\vec{p}-\vec{p''}|)\sum_{\gamma_1''\gamma_2''}\langle
f_{\vec{p}\gamma_1}|f_{\vec{p''}\gamma_1''}\rangle\langle f_{\vec{p}\gamma_2}|f_{\vec{p''}\gamma_2''}\rangle\nonumber\\
\times\Gamma_{\gamma_1''\gamma_2''\gamma_1'\gamma_2'}(\vec{p''},\vec{p'},E)\sum_{\omega_n}
G^{0(11)}_{\gamma_1''\gamma_1''}\left(\vec{p''},E/2+i\omega_n\right)\nonumber\\
\times G^{0(22)}_{\gamma_2''\gamma_2''}\left(\vec{p''},-E/2+i\omega_n\right),\nonumber\\
\label{BS1}
\end{eqnarray}
where $\Gamma_{\gamma_1\gamma_2\gamma_1'\gamma_2'}(\vec{p},\vec{p'},E)$ is the vertex part describing mutual scattering of the electron from the
electron-doped layer, located in the band $\gamma_1$ in the initial state and in the band $\gamma_1'$ in the final state, and the electron from the
hole-doped layer with the initial $-\gamma_2$ and final $-\gamma_2'$ bands; $\vec{p}$ and $\vec{p'}$ are the relative momentums of two electrons
before and after scattering respectively, $E$ is the total energy of electrons. At the onset of instability, the pole first arise at $E=0$
\cite{Abrikosov}. The frequency sums in (\ref{BS1}) at $E=0$ equal
\begin{eqnarray}
S_{\gamma_1\gamma_2}(p)=-T\sum_{\omega_n}G^{0(11)}_{\gamma_1\gamma_1}(\vec{p},i\omega_n)G^{0(22)}_{\gamma_2\gamma_2}(\vec{p},i\omega_n)\nonumber\\=
\frac{1-n_\mathrm{F}(\xi_{p\gamma_1})-n_\mathrm{F}(\xi_{p\gamma_2})}{\xi_{p\gamma_1}+\xi_{p\gamma_2}},\nonumber
\end{eqnarray}
where $n_\mathrm{F}(E)=[\exp(E/T)+1]^{-1}$ is the Fermi distribution. For certain band indices we have
\begin{eqnarray}
S_{\gamma\gamma}(p)=\frac1{2\xi_{p\gamma}}\tanh\frac{\xi_{p\gamma}}{2T},\label{Pblock1}\\
S_{\gamma,-\gamma}(p)=\frac{\sinh(\mu/T)}{4\mu\cosh(\xi_{p+}/2T)\cosh(\xi_{p-}/2T)}.\label{Pblock2}
\end{eqnarray}

We seek an instability in the $s$-wave scattering channel, so
$\Gamma_{\gamma_1\gamma_2\gamma_1'\gamma_2'}(\vec{p},\vec{p'})\equiv\Gamma_{\gamma_1\gamma_2\gamma_1'\gamma_2'}(\vec{p},\vec{p'},E=0)$ is independent
on directions of the vectors $\vec{p}$ and $\vec{p'}$, and after angular integration the equation (\ref{BS1}) takes the form:
\begin{eqnarray}
\Gamma_{\gamma_1\gamma_2\gamma_1'\gamma_2'}(p,p')=\frac{1+\gamma_1\gamma_2\gamma_1'\gamma_2'}4V_0(p,p')+
\frac{\gamma_1\gamma_1'\gamma_2\gamma_2'}4\nonumber\\
\times V_1(p,p')+\int\frac{p''\,dp''}{2\pi}\sum_{\gamma_1''\gamma_2''} \left[\frac{1+\gamma_1\gamma_2\gamma_1''\gamma_2''}4V_0(p,p'')\right.\nonumber\\
\left.+\frac{\gamma_1\gamma_1''+\gamma_2\gamma_2''}4V_1(p,p'')\right]
S_{\gamma_1''\gamma_2''}(p'')\Gamma_{\gamma_1''\gamma_2''\gamma_1'\gamma_2'}(p'',p').\nonumber\\
\label{BS2}
\end{eqnarray}
We observe immediately that 8 components $\Gamma_{\gamma_1\gamma_2\gamma_1'\gamma_2'}$, having three band indices of one sign and the last index of
the opposite sign, identically vanish. The equations (\ref{BS2}) for remaining 8 components decouple by pairs onto 4 systems. Let seek their
solutions in a form of the vertex function, constant within the pairing region:
$\Gamma_{+\gamma_2\gamma_1'\gamma_2'}(p,p')=\Gamma_{+\gamma_2\gamma_1'\gamma_2'}\Theta(w+\mu-v_\mathrm{F}p)$,
$\Gamma_{-\gamma_2\gamma_1'\gamma_2'}(p,p')=\Gamma_{-\gamma_2\gamma_1'\gamma_2'}\Theta(w-\mu-v_\mathrm{F}p)$; the pairing potentials, as in the gap
equations (\ref{Sc3}), can be replaced by their values at the Fermi level \cite{Note3}. Then the equations (\ref{BS2}) for non-zero vertex components
are
\begin{eqnarray}
\sum_\gamma\left(\delta_{\gamma_1\gamma}-\lambda_{\gamma\gamma_1}I_{\gamma\gamma}\right)\Gamma_{\gamma\gamma\gamma_2\gamma_2}=
\frac{\lambda_{\gamma_1\gamma_2}}{\mathcal{N}},\label{BS3}
\end{eqnarray}
\begin{eqnarray}
\sum_\gamma\left(\delta_{\gamma_1\gamma}-\lambda_{\gamma\gamma_1}I_{\gamma-\gamma}\right)\Gamma_{\gamma-\gamma\gamma_2-\gamma_2}=
\frac{\lambda_{\gamma_1\gamma_2}}{\mathcal{N}},\label{BS4}
\end{eqnarray}
where $\lambda_{\gamma_1\gamma_2}=\lambda_a$ at $\gamma_1=\gamma_2$ and $\lambda_b$ otherwise, and the integrals
\begin{eqnarray}
I_{\gamma_1\gamma_2}=\frac{v_\mathrm{F}}{p_\mathrm{F}}\int\limits_0^{(w+\gamma\mu)/v_\mathrm{F}}S_{\gamma_1\gamma_2}(p)p\,dp\label{I12}
\end{eqnarray}
are introduced.

The condition of a singularity of the vertex components $\Gamma_{\gamma\gamma\gamma'\gamma'}$, corresponding to a band-diagonal pairing, is
equivalent to vanishing of the determinant of the systems (\ref{BS3}) and reads
\begin{eqnarray}
(1-\lambda_aI_{++})(1-\lambda_aI_{--})-\lambda_b^2I_{++}I_{--}=0.\label{Tc_eq1}
\end{eqnarray}
Similarly, the condition of a singularity of the components $\Gamma_{\gamma-\gamma\gamma'-\gamma'}$, corresponding to a band off-diagonal pairing,
follows from (\ref{BS4}):
\begin{eqnarray}
(1-\lambda_aI_{+-})(1-\lambda_aI_{-+})-\lambda_b^2I_{+-}I_{-+}=0.\label{Tc_eq2}
\end{eqnarray}

\begin{figure}[t]
\begin{center}
\includegraphics[width=0.85\columnwidth]{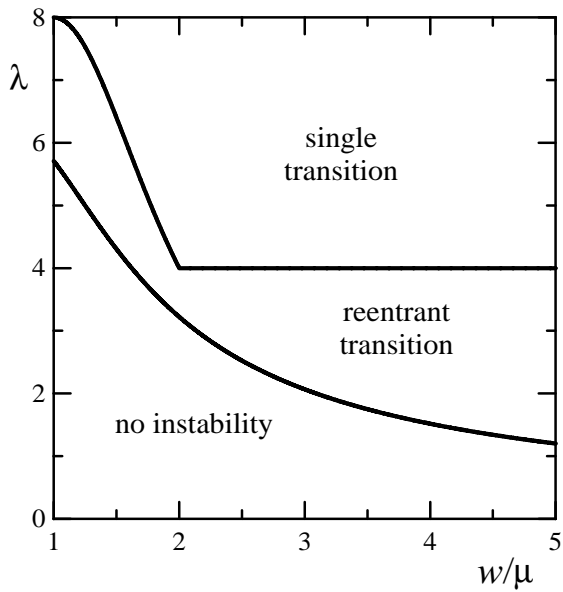}
\end{center}
\caption{\label{fig8}Phase diagram of the system with respect to a band off-diagonal pairing in the variables of the pairing region half-width $w$,
normalized on the chemical potential $\mu$, and of the dimensionless pairing constant $\lambda$. In the region of a reentrant transition, an
instability occurs only within some temperature interval.}
\end{figure}

For a qualitative analysis, assume $\lambda_a=\lambda_b=\lambda/2$, then (\ref{Tc_eq1}) is reduced to $I_{++}+I_{--}=2/\lambda$ and (\ref{Tc_eq2})
--- to $I_{+-}+I_{-+}=2/\lambda$. As seen from (\ref{I12}), the integral $I_{++}$ is logarithmically divergent when $T\rightarrow0$, whereas all the
other integrals $I_{+-}$, $I_{-+}$, $I_{--}$ converge to finite limits. This means that the instability with respect to a band-diagonal pairing
exists at arbitrarily small $\lambda$, whereas the band off-diagonal pairing instability occurs only when the coupling constant exceeds some
threshold value dependent on $w$.

Fig.~\ref{fig8} shows the phase diagram of the system with respect to the band off-diagonal pairing in the variables of pairing region half-width $w$
and coupling constant $\lambda$. At fixed $w$ and increasing $\lambda$, the stable normal state of the system changes into a state unstable in some
temperature interval from a nonzero minimum value to a maximal one. Thus, on increasing $T$, a \emph{reentrant} phase transition of the system into a
superfluid state occurs (by analogy to reentrant superconductivity). At further increase of $\lambda$, the minimal temperature restricting a region
of pairing instability turns into zero, and the system demonstrates conventional single phase transition. Such an unusual behavior, involving a
combination of single and reentrant transitions, is a consequence of unconventional interband Pauli-blocking term (\ref{Pblock2}) suppressing pairing
in some regions of phase space at finite temperatures. In contrast, the intraband Pauli-blocking terms (\ref{Pblock1}) have a conventional
hyperbolic-tangent form.

It has been shown in Sec.~\ref{part3} that the maximal value of intra- and interband coupling constants $\lambda_{a,b}$ is 1/16. At the same time,
the pairing region half-width $w$ does not exceed $8\mu\times2.19\approx17.5\mu$. The point $w/\mu=17.5$, $\lambda=1/8$ lies far in the region of
stability of the normal state in Fig.~\ref{fig8}, thus an electron-hole pairing in graphene bilayer cannot be of band off-diagonal character. Since
band-diagonal and band off-diagonal pairings are mutually concurrent, then it is the former, considered in Sec.~\ref{part3}, that should be settled
in the system at low temperatures.

Solving the equation (\ref{Tc_eq1}) numerically, one can find the critical temperature $T_\mathrm{c}$ concerning band-diagonal pairing. At
$T_\mathrm{c}\ll\mu$, $T_\mathrm{c}\ll w$ we can derive an asymptotical formula for $T_\mathrm{c}$. In this case, the integrals $I_{\gamma\gamma}$
(\ref{I12}) are
\begin{eqnarray}
I_{++}\approx\frac{w-\mu}{2\mu}+\ln\frac{2e^\gamma\sqrt{w\mu}}{\pi T},\;I_{--}\approx\frac{w-\mu}{2\mu}+\ln\sqrt{\frac\mu{w}},\nonumber
\end{eqnarray}
where $\gamma\approx0.577$ is the Euler constant. Substituting these asymptotics into (\ref{Tc_eq2}), we get the result
\begin{eqnarray}
T_\mathrm{c}=\frac{e^\gamma}\pi\,\Delta_+,\label{Tc}
\end{eqnarray}
coinciding with the usual BCS critical temperature to zero-temperature gap ratio, where $\Delta_+$ is given by (\ref{Delta2}). We have also solved
eq.~(\ref{Tc_eq1}) numerically at various $r_\mathrm{s}$ and $w$, and obtained that the ratio of $T_\mathrm{c}$ to $\Delta_+(T=0)$ is close to
(\ref{Tc}) at rather large $T_\mathrm{c}$ as well.

The Kosterlitz-Thouless superfluid transition temperature $T_\mathrm{KT}$ is determined by a condition of vortex pair dissociation
\begin{eqnarray}
\frac{\pi\rho_\mathrm{s}(T_\mathrm{KT})}2=N_\mathrm{KT}T_\mathrm{KT},\label{T_KT}
\end{eqnarray}
where $\rho_\mathrm{s}(T)$ is superfluid density (phase stiffness) at finite temperature, equal to $\mu/4\pi$ at $T=0$ and vanishing at the
mean-field transition temperature $T_\mathrm{c}$ \cite{Kharitonov2}. The factor $N_\mathrm{KT}$ takes into account vortex fractionalization; it is 1
for usual vortices and 4 for half-vortices considered in \cite{Aleiner} and having twice smaller energy then the former ones. In graphene bilayer,
\emph{quarter-vortices} are possible as topological excitations in $U(4)$ order parameter, being matrix with respect to spin projections and valleys
of paired particles (see Sec.~\ref{part2}). These quarter-vortices have four times smaller energy with respect to usual vortices, as can be shown by
analogy with \cite{Aleiner}, and so are the most favorable to be created; this leads to $N_\mathrm{KT}=4^2=16$ for our system (which differs from the
result $N_\mathrm{KT}=4$ of \cite{Kharitonov2}). Moreover, a participation of the remote bands in the pairing should lead to corrections of an
expression for $\rho_\mathrm{s}(T)$ as compared to the one-band model, although at weak coupling, when one-band regime effectively takes place,
$T_\mathrm{KT}$ differs insignificantly from the mean-field temperature $T_\mathrm{c}$. Detailed analysis of superfluid properties of graphene
bilayer in the multi-band regime at strong coupling and determination of corresponding $T_\mathrm{KT}$ will be addressed to future work.

\section{Conclusions}\label{part5}
In the present work we have considered the ultrarelativistic regime of electron-hole pairing in graphene bilayer, arising at strong coupling. The
characteristic feature of the ultrarelativistic pairing is involvement of particles from both the bands including the Fermi surfaces and the remote
bands into the pairing correlations. Such a multi-band pairing has been described in the mean-field approximation by means of the modified
diagrammatic technique. The Gor'kov equations obtained involve the gap, described by a matrix with band indices. In the band-diagonal pairing regime,
this gap is reduced to the two scalar gaps: one in the conduction band and the other, smaller by magnitude, in the valence band. A relative phase of
these gaps is fixed, while their amplitudes are different as larger, as larger the first angular harmonic of the pairing potential with respect to
the zeroth one is. At strong coupling, the magnitudes of the gaps are comparable, so a pairing within the valence band is of equal importance with
one in the conduction band. The fixation of the relative phase of the gaps becomes weak at weak coupling, and thus gapped and soliton-like
excitations, corresponding to oscillation of the relative phase, can arise.

We have studied an instability of an unpaired state of the system appearing as a singularity of the vertex part of electron-hole scattering obeying
the Bethe-Salpeter equation. A search for instability in various channels, corresponding to various band structures of the vertex part, has led to
the instability conditions for band-diagonal (where electron and hole are situated in the same band both before and after scattering) and band
off-diagonal (when the scattering participants are situated in different bands) pairings. The band off-diagonal instability exists only when the
coupling constant exceeds some threshold depending on a width of the pairing region. In some range of coupling strengths the instability occurs only
at temperatures in a range from nonzero minimal to some maximal value, however at large enough coupling constant the lower bound of the temperature
region turns into zero. Numerical estimations show that band off-diagonal instability due to screened Coulomb interaction does not realize. As for an
instability in the band-diagonal channel, it exists at arbitrarily weak coupling and corresponds to the band-dia\-gonal pairing considered in the
paper at $T=0$. Determination of Kosterlitz-Thouless transition temperature at strong coupling requires additional study (e.g., taking into account
vortex fractionalization and participation of remote bands) and will be performed in future work.

Estimations of a gap value at $T=0$ have been made in two approximations: in the approximation of constant gaps and in the approximation of separable
potential. In the first case we neglect a dependence of the gap functions on momentum and restrict momentum integration in the gap equations by some
pairing region around the Fermi surface. In the second case we handle a separable part of the statically screened interaction (which leads to
underestimation of a gap value) and do not need any momentum cutoff. The numerical estimations show that in the first case the maximum gap value is
rather small in agreement the BCS-type estimation \cite{Kharitonov1}, if the pairing region half-width equals to the characteristic frequency of
plasma oscillations in the system. Deviations from the BCS result owing to the multi-band pairing character can be significant only when the pairing
region half-width is much larger than the chemical potential (when the pairing region half-width is not large but the pairing is multi-band yet, then
the gap can be approximately twice larger than in the one-band regime, as shown in \cite{Ohsaku2}). However, calculation results in the separable
potential approximation indicate considerable deviations from the BCS result due to both a dependence of the gap function and the interaction
potential and integration on momentum in a wide region. It is shown that the separable potential approximation yields the gap values, underestimated
with respect to solving the integral gap equations with the statically screened potential; at large $r_\mathrm{s}$ the gap turns out to be comparable
to the chemical potential. This result indicates that more refined consideration of the problem is necessary in the strong coupling regime. Such a
consideration should include, in particular, taking into account retardation and damping effects in a dynamical screening of the electron-hole
interaction and going beyond the mean field approximation (i.e. taking into account vertex corrections and chemical potential renormalization), which
will be considered separately.

Nevertheless, as is well known, the dynamically scree\-ned potential is close to the statically screened one at low frequencies, tends to unscreened
Coulomb potential at high frequencies and passes through a region of repulsion concerned with plasma oscillations between these two limits. Since the
unscreened potential acts only at large momentums and frequencies far from the Fermi surface, where it differs insignificantly from the statically
screened potential, than the substitution of the overall potential by statically screened one is suitable in the regions both very close and very far
from the Fermi surface. A contribution from intermediate region where the role of plasmons is considerable should suppress pairing, and quantitative
study of this effect requires formulation and solution of Eliashberg-type equations \cite{Lozovik_Ogarkov,Eliashberg}.

The work is supported by the Russian Foundation for Basic Research and by the Program of the Russian Academy of Sciences. A.A.S. acknowledges support
given by the Dynasty Foundation and by the Russian Science Support Foundation.


\begin{thebibliography}{00}

\bibitem{Novoselov1}
K.S. Novoselov, A.K. Geim, S.V. Morozov et al., Science \textbf{306}, 666 (2004).

\bibitem{Novoselov2}
K.S. Novoselov, A.K. Geim, S.V. Morozov et al., Nature \textbf{438}, 197 (2005).

\bibitem{Geim}
A.K. Geim, K.S. Novoselov, Nature Materials \textbf{6}, 183 (2007).

\bibitem{Katsnelson1}
M.I. Katsnelson, Materials Today \textbf{10}(1--2), 20 (2007).

\bibitem{CastroNeto}
A.H. Castro Neto, F. Guinea, N.M.R. Peres, K.S.~Novoselov, A.K. Geim, Rev. Mod. Phys. \textbf{81}, 109 (2009).

\bibitem{Katsnelson2}
M.I. Katsnelson, K.S. Novoselov, Solid State Commun. 143 (2007) 3; M.I. Katsnelson, K.S. Novoselov, A.K. Geim, Nature Phys. \textbf{2}, 620 (2006).

\bibitem{Uchoa}
B. Uchoa, A.H. Castro Neto, Phys. Rev. Lett. \textbf{98}, 146801 (2007).

\bibitem{Black-Schaffer}
A.M. Black-Schaffer, S. Doniach, Phys. Rev. B \textbf{75}, 134512 (2007); C. Honerkamp, Phys. Rev. Lett. \textbf{100}, 164404 (2008).

\bibitem{Gonzalez}
J. Gonz\'{a}lez, Phys. Rev. B \textbf{78}, 205431 (2008).

\bibitem{BCS}
J. Bardeen, L.N. Cooper, J.R. Schrieffer, Phys. Rev. \textbf{108}, 1175 (1957).

\bibitem{Keldysh}
L.V. Keldysh, Yu.V. Kopaev, Sov. Phys. Solid State \textbf{6}, 2219 (1965); D. J\'{e}rome, T.M. Rice, W. Kohn, Phys. Rev. \textbf{158}, 462 (1967);
R.R. Guseinov, L.V. Keldysh, JETP \textbf{36}, 1193 (1972).

\bibitem{Lozovik_Yudson}
Yu.E. Lozovik, V.I. Yudson, JETP Lett. \textbf{22}, 274 (1975); JETP \textbf{44}, 389 (1976); Solid State Commun. \textbf{19}, 391 (1976).

\bibitem{Moskalenko}
S.A. Moskalenko, D.W. Snoke, \textit{Bose-Einstein Condensation of Excitons and Biexcitons and Coherent Nonlinear Optics with Excitons} (Cambridge
Univ. Press, Cambridge, 2000).

\bibitem{Timofeev}
V.B. Timofeev, Phys.-Usp. \textbf{48}, 295 (2005).

\bibitem{Butov}
L.V. Butov, J. Phys.: Condens. Matter \textbf{16}, R1577 (2004).

\bibitem{Eisenstein}
J.P. Eisenstein, A.H. MacDonald, Nature \textbf{432}, 691 (2004).

\bibitem{Schmidt1}
H. Schmidt, T. L\"{u}dtke, P. Barthold et al., Appl. Phys. Lett. \textbf{93}, 172108 (2008).

\bibitem{McCann}
E. McCann, V.I. Fal'ko, Phys. Rev. Lett. \textbf{96}, 086805 (2006).

\bibitem{Lozovik_Sokolik}
Yu.E. Lozovik, A.A. Sokolik, JETP Lett. \textbf{87}, 55 (2008); Yu.E.~Lozovik, S.P. Merkulova, A.A. Sokolik, Phys.-Usp. \textbf{51}, 727 (2008).

\bibitem{Nozieres}
P. Nozi\`{e}res, S. Schmitt-Rink, J. Low Temp. Phys. \textbf{59}, 195 (1985).

\bibitem{Note1}
This conclusion breaks if characteristic energy of attraction is so large that leads to inapplicability of the Dirac approximation for electron
dynamics used in the present work --- such a case was considered in \cite{Zhao}. Moreover, BCS-BEC crossover in graphene appears in the presence of
external magnetic field, where localized magnetoexcitons can form and Bose-condense \cite{Berman}, analogously to semiconductor structures
\cite{Berman2}.

\bibitem{Zhao}
E. Zhao, A. Paramekanti, Phys. Rev. Lett. \textbf{97}, 230404 (2006).

\bibitem{Berman}
O.L. Berman, Yu.E. Lozovik, G. Gumbs, Phys. Rev. B \textbf{77}, 155433 (2008).

\bibitem{Berman2}
O.L. Berman, Yu.E. Lozovik, D.W. Snoke, R.D. Coalson, Phys. Rev. B \textbf{70}, 235310 (2004); I.V. Lerner, Yu.E. Lozovik, JETP \textbf{53}, 763
(1981); A.B. Dzyubenko, Yu.E. Lozovik, J. Phys. \textbf{24}, 415 (1991).

\bibitem{Lozovik_Sokolik_JConf}
Yu.E. Lozovik, A.A. Sokolik, J. Phys.: Conf. Ser. \textbf{129}, 012003 (2008).

\bibitem{Min}
H. Min, R. Bistritzer, J.-J. Su, A.H. MacDonald, Phys. Rev. B \textbf{78}, 121401(R) (2008); R. Bistritzer, A.H. MacDonald, Phys. Rev. Lett.
\textbf{101}, 256406 (2008).

\bibitem{Zhang}
C.-H. Zhang, Y.N. Joglekar, Phys. Rev. B \textbf{77}, 233405 (2008).

\bibitem{Seradjeh}
B. Seradjeh, H. Weber, M. Franz, Phys. Rev. Lett. \textbf{101}, 264404 (2008).

\bibitem{Kharitonov1}
M.Yu. Kharitonov, K.B. Efetov, Phys. Rev. B \textbf{78}, 241401(R) (2008).

\bibitem{Bistritzer}
R. Bistritzer, H. Min, J.-J. Su, A.H. MacDonald, cond-mat/0810.0331v1.

\bibitem{Kharitonov2}
M.Yu. Kharitonov, K.B. Efetov, cond-mat/0903.4445v1.

\bibitem{Kopnin}
N.B. Kopnin, E.B. Sonin, Phys. Rev. Lett. \textbf{100}, 246808 (2008).

\bibitem{Aleiner}
I.L. Aleiner, D.E. Kharzeev, A.M. Tsvelik, Phys. Rev. B \textbf{76}, 195415 (2007).

\bibitem{Pisarski}
R.D. Pisarski, D.H. Rischke, Phys. Rev. D \textbf{60}, 094013 (1999).

\bibitem{Ohsaku1}
T. Ohsaku, Phys. Rev. B \textbf{65} 024512 (2001); \textbf{66}, 054518 (2002).

\bibitem{Ohsaku2}
T. Ohsaku, Int. J. Mod. Phys. B \textbf{18}, 1771 (2004).

\bibitem{Lozovik_Ogarkov}
Yu.E. Lozovik, S.L. Ogarkov, A.A. Sokolik, JETP, to be published.

\bibitem{Semenoff}
G.W. Semenoff, Phys. Rev. Lett. \textbf{53}, 2449 (1984).

\bibitem{Akhiezer}
A.I. Akhiezer, V.B. Berestetskii, \textit{Quantum Electrodynamics} (Wiley, New York, 1965).

\bibitem{Abrikosov}
A.A. Abrikosov, L.P. Gor'kov, I.E. Dzyaloshinskii, \textit{Methods of Quantum Field Theory in Statistical Physics} (Dover, New York, 1963).

\bibitem{Gor'kov}
L P. Gor'kov, JETP \textbf{7}, 505 (1958).

\bibitem{Sigrist}
M. Sigrist, K. Ueda, Rev. Mod. Phys. \textbf{63}, 239 (1991).

\bibitem{Apenko}
S.M. Apenko, D.A. Kirzhnits, Yu.E. Lozovik, Phys. Lett. A \textbf{92}, 107 (1982).

\bibitem{DasSarma_Madhukar}
S. Das Sarma, A. Madhukar, Phys. Rev. B \textbf{23}, 805 (1981).

\bibitem{DePalo}
S. De Palo, F. Rapisarda, G. Senatore, Phys. Rev. Lett. \textbf{88}, 206401 (2002); Y.N. Joglekar, A.V. Balatsky, S. Das Sarma, Phys. Rev. B
\textbf{74}, 233302 (2006).

\bibitem{Lozovik_Kurbakov}
Yu.E. Lozovik, I.L. Kurbakov, G.E. Astrakharchik, M. Willander, JETP \textbf{106}, 296 (2008); Yu.E. Lozovik, I.L.~Kurbakov, M.~Willander, Phys.
Lett. A \textbf{366}, 487 (2007).

\bibitem{Wuncsh}
B. Wunsch, T. Stauber, F. Sols, F. Guinea, New J. Phys. \textbf{8}, 318 (2006).

\bibitem{Hwang}
E.H. Hwang, S. Das Sarma, Phys. Rev. B \textbf{75}, 205418 (2007).

\bibitem{Note2}
Note that, unlike the pairing of massless fermions coupled to scalar bosons considered in \cite{Pisarski}, in our case the Coulomb interaction is
longitudinally-vectorial. Under scalar interaction, the system of gap equations differs from (\ref{Sc3}) by the permutation $V_a\leftrightarrow V_b$,
and consequently the ratio $\Delta_-/\Delta_+$ is qualitatively different from the presented results \cite{Pisarski}: $\Delta_-$ is always larger
than $\Delta_+$, and at weak coupling $\Delta_-/\Delta_+$ tends to infinity.

\bibitem{Milovanovic}
M.V. Milovanovi\'{c}, Phys. Rev. B \textbf{78}, 245424 (2008).

\bibitem{Lozovik_Poushnov}
Yu.E. Lozovik, A.V. Poushnov, Phys. Lett. A \textbf{228}, 399 (1997).

\bibitem{Khodel}
V.A. Khodel, V.V. Khodel, J.W. Clark, Nucl. Phys. A \textbf{598}, 390 (1996).

\bibitem{Kosterlitz}
J.M. Kosterlitz, D.J. Thouless, J. Phys. C \textbf{6}, 1181 (1973).

\bibitem{Note3}
Strictly speaking, using zero-temperature expressions from Sec.~\ref{part3} for a pairing potential is valid if an influence of a finite temperature
on the polarization operator is negligible on the characteristic momentum $p_\mathrm{F}/r_\mathrm{s}$, i.e. when $T_\mathrm{c}\ll\mu/r_\mathrm{s}$.

\bibitem{Eliashberg}
G.M. Eliashberg, JETP \textbf{11}, 696 (1960).

\end{thebibliography}
\end{document}